\documentclass[prd,unsortedaddress,showpacs,amssymb,amsmath,twocolumn,superscriptaddress]{revtex4}
\usepackage{epsf,graphicx}
\usepackage{epstopdf}
\begin{document}
\newcommand{\deltl}{$\Delta L = 2$}
\newcommand{\pmmdec}{$\Xi^- \to p \mu^- \mu^-$}
\newcommand{\smgauge}{$SU(2)_L \times U(1)_Y$}
\newcommand{\casdec}{$\Xi^- \to \Lambda \pi^-$}
\newcommand{\lamdec}{$\Lambda \to p \pi^-$}
\newcommand{\kdec}{$K^- \to \pi^-\pi^-\pi^+$}
\newcommand{\ktpi}{$K^{\pm} \to \pi^{\mp} \pi^{\mp} \pi^{\pm}$}
\title {Measurement of the $\alpha$ Asymmetry Parameter for the $\Omega^-\to\Lambda K^-$ Decay}

\affiliation{Institute of Physics, Academia Sinica, Taipei 11529,
   Taiwan, Republic of China}
\affiliation{University of California, Berkeley, California 94720, USA}
\affiliation{Fermi National Accelerator Laboratory, Batavia, Illinois 60510, USA}
\affiliation{Universidad de Guanajuato, 37000 Le\'{o}n, Mexico}
\affiliation{Illinois Institute of Technology, Chicago, Illinois 60616, USA}
\affiliation{Universit\'{e} de Lausanne, CH-1015 Lausanne, Switzerland}
\affiliation{Lawrence Berkeley National Laboratory, Berkeley, California 94720, USA}
\affiliation{University of Michigan, Ann Arbor, Michigan 48109, USA}
\affiliation{University of South Alabama, Mobile, Alabama 36688, USA}
\affiliation{University of Virginia, Charlottesville, Virginia 22904, USA}

\author{Y. C. Chen}
\affiliation{Institute of Physics, Academia Sinica, Taipei 11529, Taiwan,
   Republic of China}
\author{R. A. Burnstein}
\affiliation{Illinois Institute of Technology, Chicago, Illinois 60616, USA}
\author{A. Chakravorty}
\affiliation{Illinois Institute of Technology, Chicago, Illinois 60616, USA}
\author{A. Chan}
\affiliation{Institute of Physics, Academia Sinica, Taipei 11529, Taiwan,
   Republic of China}
\author{W. S. Choong}
\affiliation{University of California, Berkeley, California 94720, USA}
\affiliation{Lawrence Berkeley National Laboratory, Berkeley, California
   94720, USA}
\author{K. Clark}
\affiliation{University of South Alabama, Mobile, Alabama 36688, USA}
\author{E. C. Dukes}
\affiliation{University of Virginia, Charlottesville, Virginia 22904, USA}
\author{C. Durandet}
\affiliation{University of Virginia, Charlottesville, Virginia 22904, USA}
\author{J. Felix}
\affiliation{Universidad de Guanajuato, 37000 Le\'{o}n, Mexico}
\author{G. Gidal}
\affiliation{Lawrence Berkeley National Laboratory, Berkeley, California
   94720, USA}
\author{P. Gu}
\affiliation{Lawrence Berkeley National Laboratory, Berkeley, California
   94720, USA}
\author{H. R. Gustafson}
\affiliation{University of Michigan, Ann Arbor, Michigan 48109, USA}
\author{C. Ho}
\affiliation{Institute of Physics, Academia Sinica, Taipei 11529, Taiwan,
   Republic of China}
\author{T. Holmstrom}
\affiliation{University of Virginia, Charlottesville, Virginia 22904, USA}
\author{M. Huang}
\affiliation{University of Virginia, Charlottesville, Virginia 22904, USA}
\author{C. James}
\affiliation{Fermi National Accelerator Laboratory, Batavia, Illinois 60510, USA}
\author{C. M. Jenkins}
\affiliation{University of South Alabama, Mobile, Alabama 36688, USA}
\author{D. M. Kaplan}
\affiliation{Illinois Institute of Technology, Chicago, Illinois 60616, USA}
\author{L. M. Lederman}
\affiliation{Illinois Institute of Technology, Chicago, Illinois 60616, USA}
\author{N. Leros}
\affiliation{Universit\'{e} de Lausanne, CH-1015 Lausanne, Switzerland}
\author{M. J. Longo}
\affiliation{University of Michigan, Ann Arbor, Michigan 48109, USA}
\author{F. Lopez}
\affiliation{University of Michigan, Ann Arbor, Michigan 48109, USA}
\author{L. C. Lu}
\affiliation{University of Virginia, Charlottesville, Virginia 22904, USA}
\author{W. Luebke}
\affiliation{Illinois Institute of Technology, Chicago, Illinois 60616, USA}
\author{K. B. Luk}
\affiliation{University of California, Berkeley, California 94720, USA}
\affiliation{Lawrence Berkeley National Laboratory, Berkeley, California
   94720, USA}
\author{K. S. Nelson}
\affiliation{University of Virginia, Charlottesville, Virginia 22904, USA}
\author{H. K. Park}
\affiliation{University of Michigan, Ann Arbor, Michigan 48109, USA}
\author{J.--P. Perroud}
\affiliation{Universit\'{e} de Lausanne, CH-1015 Lausanne, Switzerland}
\author{D. Rajaram}
\affiliation{Illinois Institute of Technology, Chicago, Illinois 60616, USA}
\author{H. A. Rubin}
\affiliation{Illinois Institute of Technology, Chicago, Illinois 60616, USA}
\author{P. K. Teng}
\affiliation{Institute of Physics, Academia Sinica, Taipei 11529, Taiwan,
   Republic of China}
\author{J. Volk}
\affiliation{Fermi National Accelerator Laboratory, Batavia, Illinois 60510, USA}
\author{C. G. White}
\affiliation{Illinois Institute of Technology, Chicago, Illinois 60616, USA}
\author{S. L. White}
\affiliation{Illinois Institute of Technology, Chicago, Illinois 60616, USA}
\author{P. Zyla}
\affiliation{Lawrence Berkeley National Laboratory, Berkeley, California
   94720, USA}

\collaboration{The HyperCP Collaboration}
\noaffiliation

\date{\today}

\begin{abstract}
We have measured the $\alpha$ parameter of the $\Omega^-\to\Lambda K^-$ decay using data collected with the HyperCP spectrometer during the 1997 fixed-target run at Fermilab. Analyzing a sample of 0.96\,million  $\Omega^-\to\Lambda K^-,$ $\Lambda\to p\pi^-$ decays, we obtain $\alpha_\Omega\alpha_\Lambda = [1.33\pm0.33\,(\rm stat)\pm0.52\,(\rm syst)]\times 10^{-2}$. With the accepted value of $\alpha_\Lambda$, $\alpha_\Omega$ is found to be $[2.07\pm0.51\,(\rm stat)\pm0.81\,(\rm syst)]\times 10^{-2}$. 
\end{abstract}

\pacs{11.30.Er, 13.30.Eg, 14.20.Jn}

\maketitle

\section{Introduction}
The $\Omega^-$ has played a celebrated role in particle physics.  
Its discovery~\cite{Barnes} in 1964 confirmed its prediction~\cite{Gell-Mann}
as the missing member of the spin-$\frac{3} {2}$ baryon decuplet. 
However, the prediction of its spin has not yet been unambiguously verified: 
we can only say with certainty that it is not spin-$\frac{1} {2}$~\cite{Omega-spin}. If we assume that the 
$\Omega^-$ is spin-$\frac{3} {2}$ (as we shall throughout this paper), parity violation  
in weak interactions allows the final state in the $\Omega^-\to\Lambda K^-$ decay to contain 
a mixture of $P$ and $D$ waves. The relative admixture of these  
angular-momentum states is poorly known. Previous experiments have established that $\alpha_\Omega$, defined as  
\begin{equation}
\alpha_\Omega=\frac{2{\rm Re}(P^*D)}{|P|^2+|D|^2}\,,\label{eq1}
\end{equation}
is small, if not zero, the world average being  
$(-2.6\pm2.3)\times 10^{-2}$~\cite{PDG}. Theory predicts the 
$\Omega^-\to\Lambda K^-$ decay to be predominately parity conserving~\cite{Omega-theory} and hence 
dominated by the $P$-wave final state, consistent with this small value.  

\begin{figure} [htbp]
\includegraphics[width=6cm]{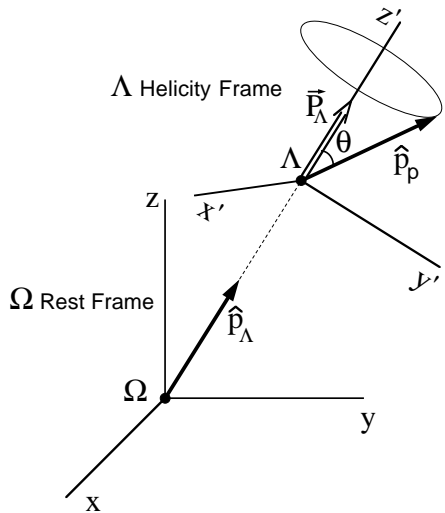}
\caption{Definition of the Lambda Helicity Frame.}\label{fig1}
\end{figure}

In the decay of an unpolarized $\Omega^-$ hyperon, the daughter $\Lambda$ is produced with a longitudinal polarization~\cite{Luk},
which leads to an asymmetry in the $\Lambda\to p\pi^-$ decay. In this case the angular distribution of the proton is 
\begin{equation}
\frac{dN} {d\cos\theta} = \frac{N_0} {2} (1 + \alpha_\Omega\alpha_\Lambda\cos\theta)\,,
\end{equation}
where $N_0$ is the total number of events, $\alpha_\Lambda$ is the decay parameter of the $\Lambda\to p\pi^-$ decay, and $\theta$ is the polar angle of the proton momentum in that $\Lambda$ rest frame whose polar axis is the direction of the $\Lambda$ momentum in the 
$\Omega^-$ rest frame (the ``Lambda Helicity Frame" shown in Fig.~\ref{fig1}). Since the polar ($z^\prime$) axis in the Lambda Helicity Frame changes direction from event to event, there is little correlation between $\theta$ and any particular region in the laboratory frame, greatly reducing many sources of bias. 

We here report a precise determination of $\alpha_\Omega$ for the
$\Omega^-\to\Lambda K^-$ decay with a sample of 0.96 million unpolarized 
$\Omega^-\to\Lambda K^-$ events which is some two orders of 
magnitude larger than those of  previous experiments. 

\section{THE HYPERCP EXPERIMENT}
\begin{figure} [htbp]
\vspace{-0.75cm}
\includegraphics[width=8.5cm]{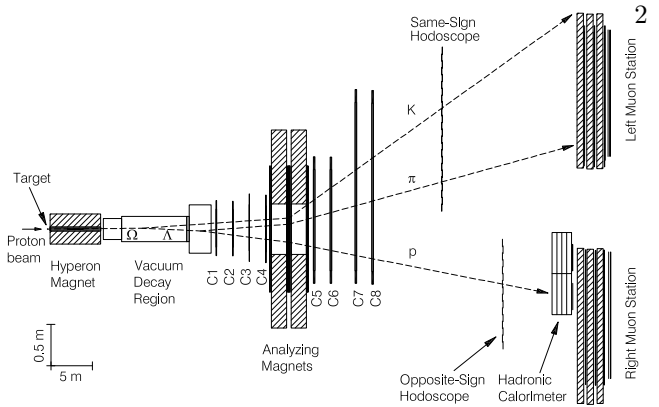}
\caption{Plan view of the HyperCP spectrometer.}\label{fig2}
\end{figure}

The data used in this analysis were taken in the 1997 fixed-target run at Fermilab using the HyperCP spectrometer~\cite{Burnstein} (Fig.~\ref{fig2}). This was a high-rate spectrometer designed to perform a sensitive search for
{\em CP} violation in charged-$\Xi$ and $\Lambda$ decays, as well as  
searches for rare or forbidden hyperon and kaon decays. The $\Omega^-$ hyperons were produced as follows. An 800\,GeV/$c$ proton beam, incident at zero degrees, was steered onto a 6-cm-long, $0.2\times0.2$\,cm$^2$ copper target situated in front of a 6.096-m-long curved collimator located within a dipole magnet (the Hyperon Magnet of Fig.~\ref{fig2}) producing a 1.667\,T field. Typically  
$7.2\times10^{9}$ protons per second were delivered onto the target in 19\,s spills occurring once per minute; this produced a 13\,MHz secondary beam exiting the collimator. The zero-degree  
incident angle dictated, through parity conservation in  
the strong interaction, that the 
$\Omega^-$'s were produced unpolarized. 

The Hyperon Magnet deflected negatively 
charged particles at the nominal secondary-beam momentum upward at a 19.5\,mrad angle; here ``up" is the $+y$ direction, the charged secondary beam moves in the $+z$ direction, and ``beam-left" is the $+x$ direction, so that the $x$, $y$, and $z$ axes form a right-handed coordinate system. The defining apertures of the collimator  
limited the momentum range of the secondary beam to  
about $120-220$\,GeV/$c$, with an average momentum of  
about 160\,GeV/$c$. Immediately following the Hyperon  
Magnet was a 13-m-long evacuated pipe (Vacuum Decay Region), which defined  
the allowed decay region of the $\Omega^-$. 

Following the Vacuum Decay Region was a magnetic spectrometer used to measure the  
momenta of the $p$, $K^-$, and $\pi^-$ from the 
$\Omega^-$ and $\Lambda$ decays. It was composed of high-rate, narrow-pitch, multiwire  proportional chambers (MWPCs), four (C1--C4) located in front of the Analyzing Magnets 
and four behind (C5--C8). Each  
chamber had four anode planes, two with vertical wires  
($X$ view) and two with wires angled at $\pm26.57^\circ$ to the vertical ($U$ and $V$ views). The wire pitch increased with distance from the target, with C1 and C2 at 1\,mm, C3  
and C4 at 1.25\,mm, C5 and C6 at 1.5\,mm, and C7 and  
C8 at 2.0\,mm. The Analyzing Magnets were two dipole  
magnets placed back to back. They had a combined field  
integral of 4.73\,T\,m. Negatively charged particles were  
deflected in the $+x$ direction.

Toward the rear of the spectrometer were the trigger  
elements. These included two scintillation-counter hodoscopes,  
both positioned outside of the envelope of the intense  
secondary beam emanating from the collimator. One (Same-Sign Hodoscope) was on the  
side of the spectrometer to which particles with charge of the same (SS) sign as the secondary beam were deflected, and the other (Opposite-Sign Hodoscope) was on the side  
to which particles of the opposite charge sign (OS) were deflected. The SS Hodoscope  
was used to detect the presence of the kaon or  
pion from the $\Omega^-\to\Lambda K^-$, $\Lambda\to p\pi^-$
decay sequence, while the OS Hodoscope was used to detect the proton from the $\Lambda$
decay. Behind the OS Hodoscope was  
the Hadronic Calorimeter, used to measure the energy of the  
proton. The trigger used to select events for this analysis (called the CAS trigger) was designed to pick out  
candidate $\Lambda\to p\pi^-$ decays and required the coincidence of at least one charged particle in each of the SS and  
OS Hodoscopes and a minimum energy of approximately  
40\,GeV in the Hadronic Calorimeter.  
(The muon detector system  
at the rear of the apparatus was not used in this analysis.)

\section{Data Analysis}
The 39 billion CAS triggers recorded in the 1997 run were reconstructed with a program 
that found all of the tracks in the event. Events with fewer
than three good tracks outside of the secondary-beam  
envelope, as well as those  
with no opposite-sign track, were discarded. For each remaining event, every possible combination of SS and OS  
tracks was used to form a $p\pi^-$ invariant mass, with  
the assumption that the SS track was a pion and the  
OS track a proton. 
The SS--OS track pair with $p\pi^-$ mass closest to the $\Lambda$ mass was  
tagged.
The tagged pair was then  
combined with the remaining SS tracks, and the three-track invariant mass was determined assuming a $p\pi^- K^-$ hypothesis. The combination with $p\pi^- K^-$ mass closest to  
the $\Omega^-$ mass was tagged. A geometric fit of the three tagged tracks then determined whether their topology was consistent with a two-vertex, three-track hypothesis, with the proton and pion tracks forming one of the vertices. 

Tight event-selection cuts were applied to produce as clean an $\Omega^-$ sample as possible. These cuts required (1) that the $\chi^2$ per degree of freedom of the geometric fit be less than 2.5; (2) that the $p\pi^-$ invariant mass be consistent with that of a $\Lambda$ ($1.1124 < m_{p\pi} < 1.1196$\,GeV/$c^2$); (3)  that the three tracks not have an invariant mass consistent with a $\Xi^-\to\Lambda\pi^-$ decay ($m_{p\pi\pi} > 1.335$\,GeV/$c^2$); (4) that the $\pi^+\pi^-\pi^-$ invariant mass be less than 0.48\,GeV/$c^2$ or greater than 0.51\,GeV/$c^2$ to remove potential $K^-\to\pi^+\pi^-\pi^-$ contamination; (5) that the $\Lambda$ ($\Omega^-$) decay vertex be at least 0.40\,m (0.60\,m) and less than 13\,m downstream of the exit of the collimator; (6) that the reconstructed $\Omega^-$ track at the exit of the collimator be at most 8.0\,mm (5.5\,mm) from the center of the collimator exit aperture in $x$ ($y$); and (7) that the $\Omega^-$ track extrapolated back to the target be at most 2.5\,mm (3\,mm) from the target center in $x$ ($y$). Loose cuts on the momenta of the various particles were also applied (with one exception described next). In addition to these cuts two more cuts were applied to improve the agreement between the data and the Hybrid Monte Carlo simulation (described below) by eliminating events populating regions in which the Hybrid Monte Carlo distributions matched the data poorly. These required that the separation of the $p$ and $\pi^-$ tracks at chamber C4 exceed 2.0\,cm in each of the three views, and that the $\pi^-$ momentum exceed 21.5\,GeV/$c$. 

\begin{figure} [htbp]
\includegraphics[width=11cm]{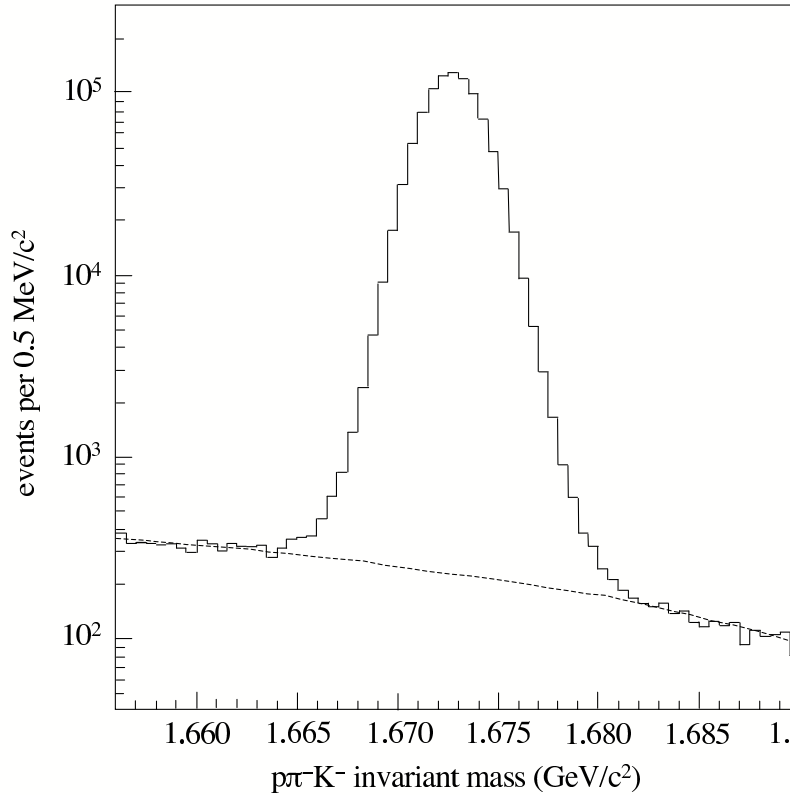}
\caption{Distribution of the $p\pi^- K^-$ invariant mass with 3rd-order-polynomial background fit superimposed.}\label{fig3}
\end{figure}

The $p\pi^- K^-$ invariant-mass distribution of the events passing these cuts is shown in Fig.~\ref{fig3} with a third-order polynomial fit to the background superimposed. The mass resolution, $\sigma \approx 1.6$\,MeV/$c^2$, is consistent with that obtained with Monte Carlo simulation. A total of 0.96\,million events lie between 1.6647 and 1.6807 GeV/$c^2$, that is, within $\pm5\,\sigma$ of the central value of the $p\pi^- K^-$ mass (1.6728\,GeV/$c^2$). The fraction of background within this mass range is 0.76\%. 

\section{Extraction of $\alpha_\Omega$}
In order to extract $\alpha_\Omega$, the slope of the proton $\cos\theta$ distribution, $s_m$, in the Lambda Helicity Frame was measured and that of the $\cos\theta$ distribution of the background, $s_b$, subtracted to find $\alpha_\Omega\alpha_\Lambda$. The effect of the acceptance on the $\cos\theta$ distribution was accounted for using a Hybrid Monte Carlo (HMC) technique~\cite{HMC} as follows. Each real candidate $\Omega^-$ event was read in by the HMC program, and the $\cos\theta$ value of its proton was determined and stored. Monte Carlo events were then generated by taking all of the parameters from the real event except the momenta of the proton and pion. New proton and pion momenta were generated in the Lambda Helicity Frame with an isotropic distribution and then Lorentz-boosted into the laboratory frame. The HMC-generated proton and pion tracks were then traced through a software model of the spectrometer and required to pass the same trigger and event-topology requirements as the real data. For every real event ten accepted HMC events were required. If after two hundred tries, ten HMC events were not accepted (a rarity given the large acceptance of the spectrometer), the real and HMC events were discarded. 

Accepted HMC events were weighted by the function 
\begin{eqnarray}
W(s_m, \cos\theta^f_i )= \frac{1 + s_m \cos\theta^f_i}{1 + s_m \cos\theta^r}&\\
\approx (1 + s_m \cos\theta^f_i )[1 - s_m\cos\theta^r &\!\!\!+ (s_m \cos\theta^r)^2 -\cdot\cdot\cdot],\label{Eq3}
\end{eqnarray} 
where $\theta^r$ is the polar angle of the real proton and $\theta^f_i$ are those of the associated HMC protons. As indicated in Eq.~\ref{Eq3}, the series expansion of the weight function in powers of $s_m \cos\theta^r$ allowed it to be written as a third-degree polynomial in the unknown slope $s_m$, with the coefficients of the polynomial stored for each bin of HMC-proton $\cos\theta^f_i$. After processing all of the real $\Omega^-$ events in this manner, a $\chi^2$ comparison between the real-proton and HMC-proton $\cos\theta$ distributions was performed and the value of the slope $s_m$ that minimized the $\chi^2$ was found. This value was corrected for background as described below to give the final value of $\alpha_\Omega\alpha_\Lambda$.
  
The analysis procedure was extensively tested on Monte Carlo events. Seven Monte Carlo samples were generated, each with one million accepted events, with $\alpha_\Omega$ values of $\pm0.2$, $\pm0.1$, $\pm0.02$, and 0.0. The average difference between the input and HMC-determined $\alpha_\Omega\alpha_\Lambda$ values was found to be $-0.0029\pm0.0014$; this offset was accounted for when the final result was extracted. 
The difference was independent of the input value.
  
\begin{figure} [htbp]
\includegraphics[width=11cm,height=9.5cm]{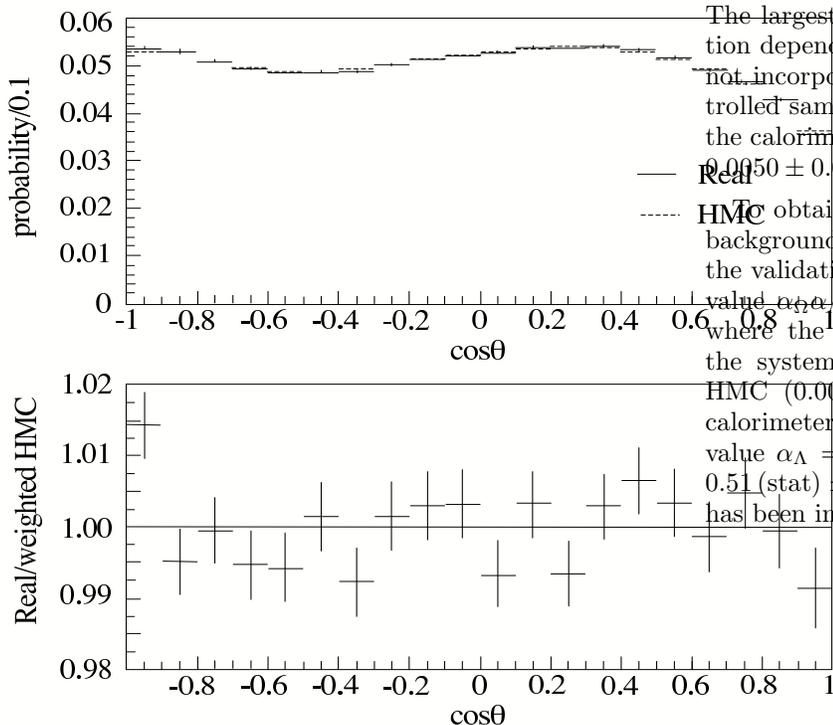}
\caption{Comparison of the real-proton and weighted-HMC-proton $\cos\theta$ distributions in the Lambda Helicity Frame, where the number of events in each plot has been normalized to one (top). The bottom plot is the ratio of the real-proton to weighted-HMC-proton $\cos\theta$ distributions. }\label{fig4}
\end{figure}

For the selected sample of 0.96 million $\Omega^-\to\Lambda K^-$ events, the extracted value of the slope of the $\cos\theta$ distribution was $s_m = 0.0115\pm0.0033$, where the error is statistical and includes the HMC statistical error. The HMC $\cos\theta$ distribution of the proton 
weighted by the best-fit value of $s_m$ and the distribution of the real data are shown in Fig.~\ref{fig4}. The $\chi^2$ of the HMC fit was 30.2 for 19 degrees of freedom. To extract $\alpha_\Omega\alpha_\Lambda$, the contribution of the background under the $\Omega^-$ mass peak was subtracted as follows. Sidebands outside of the $p\pi^- K^-$ mass range used to measure $s_m$ were analyzed as described above to determine the slope
of the $\cos\theta$ distribution of their protons, $s_b$. The lower sideband range was $1.6519 < m_{p\pi K} < 1.6647$\,GeV/$c^2$ and the upper sideband range was $1.6807 < m_{p\pi K} < 1.6935$\,GeV/$c^2$. Each sideband was subdivided into three bins in $p\pi^- K^-$ mass. The extracted slopes of the proton $\cos\theta$ distributions for the sideband data sets were consistent with each other. A HMC fit to all of the background samples gave $s_b = 0.159\pm0.038$, considerably larger than that found in the signal region. Using $s_b$ thus determined as the slope of the proton $\cos\theta$ distribution for the background events under the $\Omega^-$ mass peak, $\alpha_\Omega\alpha_\Lambda$ was extracted from $s_m$ to give $\alpha_\Omega\alpha_\Lambda = 0.0104\pm0.0033$, where the error is statistical.  

The stability of the result was studied as a function of several parameters. A run-by-run determination of $\alpha_\Omega\alpha_\Lambda$ found no significant variation. The value of $\alpha_\Omega\alpha_\Lambda$ was also measured as a function of the momentum of the $\Omega^-$ and the position of the $\Omega^-$ decay vertex. No significant dependences were found. 

\section{Systematic Studies and Corrections}
We have investigated the potential systematic uncertainties related to event selection, performance of the spectrometer, and modeling of the background under the $\Omega^-$ mass peak. The extracted value of $\alpha_\Omega\alpha_\Lambda$ was insensitive to variations in the minimum-$\pi^-$-momentum requirement by up to $\pm0.4$\,GeV/$c$, and on the separation between the $p$ and $\pi^-$ tracks at C4 by up to $\pm0.4$\,cm in each view. The uncertainty in the amount of background under the $\Omega^-$ peak within the $p\pi^- K^-$ mass window was estimated by performing first-degree and second-degree polynomial fits to the background. Since the background contribution is small, this variation caused no observable change in the extracted value of $\alpha_\Omega\alpha_\Lambda$. The uncertainty in $s_b$ led to a systematic error of 0.0003 in $\alpha_\Omega\alpha_\Lambda$. The largest systematic uncertainty came from the position dependence of the calorimeter efficiency, which was not incorporated into the HMC code. By studying controlled samples of Monte Carlo events with and without the calorimeter efficiency, we found a systematic error of $0.0050\pm0.0020$ in $\alpha_\Omega\alpha_\Lambda$. 

To obtain the final result in $\alpha_\Omega\alpha_\Lambda$ we corrected the background-subtracted value with the offset observed in the validation of the HMC procedure and arrived at the value $\alpha_\Omega\alpha_\Lambda = [1.33\pm0.33\,({\rm stat})\pm0.52\,({\rm syst})]\times 10^{-2}$, where the systematic error is the quadrature sum of the systematic uncertainties due to validation of the HMC (0.0014), background subtraction (0.0003), and calorimeter efficiency (0.0050). Using the world-average value $\alpha_\Lambda = 0.642 \pm 0.013$~\cite{PDG}, we obtain $\alpha_\Omega = [2.07\pm0.51\,({\rm stat})\pm0.81\,({\rm syst})]\times10^{-2}$, where the error in $\alpha_\Lambda$ has been included in the systematic uncertainty. 

\section{Conclusion}
In conclusion, based on a sample of 0.96\,million events from the 1997 run of HyperCP, we have obtained a precise value of the $\alpha$ parameter for the $\Omega^-\to\Lambda K^-$ decay mode. 
It is a factor of 2.4 better than the world average and may differ in sign. 
Our result also indicates that, for this decay mode, $\alpha_\Omega$ is small but, at about 
2.2-standard-deviation significance, is likely nonzero. Thus the $\Omega^-\to\Lambda K^-$ decay is predominantly parity conserving as theoretically expected. 

\begin{acknowledgments}
We are indebted to the staffs of Fermilab and the collaborating institutions for their hard work and dedication. This work was supported by the U.S. Dept.\ of Energy and the National Science Council of Taiwan, R.O.C. D.M.K. acknowledges support from the Particle Physics and Astronomy Research Council of the U.K. and the hospitality of Imperial College London while this paper was in preparation. E.C.D. and K.S.N. were partially supported by the Institute for Nuclear and Particle Physics of the University of Virginia. K.B.L. was partially supported by the Miller Institute for Basic Research in Science. 

\end{acknowledgments}

\end{document}